\documentclass[aps,10pt,prb,superscriptaddress,twocolumn,showpacs]{revtex4}
\usepackage[dvips]{graphicx}
\usepackage{color}
\usepackage{booktabs}
\usepackage{dcolumn,bm}

\begin{document}

\title{Quasi-energy spectra of graphene dots under intense ac field: field anisotropy and photon dressed quantum rings}
\author{P.~H.~Rivera}
\affiliation{Facultad de Ciencias F\'{\i}sicas, Universidad Nacional Mayor de San Marcos, Lima, Per\'u}
\author{A.~L.~C.~Pereira}
\author{P.~A.~Schulz}
\affiliation{Instituto de F\'{\i}sica ``Gleb Wathagin", Universidade Estadual de Campinas, Campinas, S\~ao Paulo, Brazil}
\date{\today}

\begin{abstract}
A graphene quantum dot under intense ac field and static low magnetic field is investigated. From a tight-binding perspective, applying a Fourier-Floquet transformation and renormalization process, we observe that graphene -intrinsically anisotropic- reveals field polarization signatures in the quasi-density of states. For the ac field polarized along the armchair direction, the dressed electronic structure shows an emergent property: an ac field induced quantum ring. This is inferred by the orientation-dependent formation of a miniband of energy states periodically modulated with increasing magnetic field, exactly analogous to the behavior of a quantum ring spectrum.  
\end{abstract}


\maketitle

\section{Introduction}

An intense research effort has been launched after the stabilization of graphene \cite{novo1}, a true two-dimensional system, due to the emergence of unconventional electronic properties with broad potential applications \cite{geim,castro}. A benchmark for a two dimensional system is the quantum Hall effect, which indeed was observed in graphene\cite{novo2,kim}, even at room temperatures\cite{novo3}.
The research effort on graphene also includes its response to intense ac electric fields\cite{gusynin,mikha,naumis,aoki}. The concurrence of an ac electric field in plane and a perpendicularly applied static magnetic field, acting simultaneously on a graphene system, will be the focus of our work here, motivated by another recently discovered emergent effect, namely the microwave induced zero resistance states in very high mobility, but still ``conventional'', quantum Hall systems\cite{mani,zudov}. One characteristic feature of these states is that they are immune to the in plane ac field polarization\cite{mani2,smet}, an aspect that continues to be addressed  theoretically in ``conventional'' 2D systems\cite{wang,inarrea}. Furthermore, the microwave intensities in these experiments could go beyond a perturbative regime, leading to the necessity of treating the electron-ac field interaction from a non-perturbative approach\cite{pablo2004}.

These elements settle the scenario and context of our work: we investigate the evolution of the graphene quasi-energy spectrum in the presence of both, an ac field parallel and a static magnetic field perpendicular to the carbon atoms plane. On the contrary to electrons at the bottom of the GaAs conduction band, graphene is intrinsically anisotropic and the response to ac fields should reveal fingerprints of this anisotropy. We address the problem from a tight-binding approach applying a Floquet-Fourier transformation and renormalization procedure that permits us to find the density of states of the quasi energy spectrum\cite{pablo2004}. This approach leads to the limitation of handling the problem only for finite systems. On the other hand, emerging effects for dot like structures may be identified, and indeed, small quantum dots out of a graphene layer is rapidly becoming a reality\cite{ensslin,novo4}. 

\section{The model}

The tight-binding approach for the graphene electronic structure coupled to an ac field is given by an Hamiltonian divided in two parts, $ H= H_o+H_i$, where $H_i$ includes the interaction of the lattice with the ac field, while the interaction of the static magnetic field with the lattice is included in $H_o$,

\begin{displaymath}
H_o=\sum_{l_1,l_2}|l_1,l_2\rangle\mathcal{E}_{l_1,l_2}\langle l_1,l_2|+
\end{displaymath} 
\begin{displaymath}
 \sum_{l_1,l_2}\Bigl\{\Bigl[|l_1,l_2\rangle{V_x\over2}\langle l_1+1,l_2|+|l_1+1,l_2\rangle{V_x\over2}\langle l_1,l_2|\Bigr]\bigr.+
\end{displaymath}
\begin{equation}
\Bigl.\Bigl[e^{i2\pi\alpha
l_1}\Bigl(|l_1,l_2\rangle{V_y\over2}\langle l_1,l_2+1|+|l_1,l_2+1\rangle{V_y\over2}\langle l_1,l_2|\Bigr)\Bigr]\Bigr\} \label{Eq.1}
\end{equation}
and
\begin{equation} H_i=e
\left(
\begin{array}{c}
a_x \\ a_y
\end{array}
\right)
F\cos \omega t\sum_{l_1,l_2}|l_1,l_2\rangle 
\left(
\begin{array}{c}
l_1 \\ l_2
\end{array}
\right)\langle l_1,l_2|, \label{Eq.2}
\end{equation}
where the index $x$ is for the ac electric field component parallel to the zigzag direction, and $y$ is
for the ac field parallel to the armchair direction, as visualized in the bricklayer representation of a graphene finite lattice, Fig.~\ref{fig.1}. Furthermore, $(l_1,l_2)$ are the index of the $(x,y)$ coordinates of the sites. The phase factor is defined as $\alpha=\Phi/\Phi_e$, where $\Phi_e=h/e$ is the quantum magnetic flux and $\Phi=a_xa_yB$ is the magnetic flux per a half
unit cell of the graphene lattice, because $a_x=\sqrt{3}a/2$, in mean value $a_y=3a/2$,  and $a=1.42$ \AA, the graphene lattice parameter. 
While the $p_z$ orbital energy will be taken constant, and we choose $\mathcal{E}_{l_1,l_2}=0$ with no diagonal disorder. The hopping parameter are $V_x$ and $V_y$ as the nearest neighbors interaction and are taken constant and equal to 2.97 eV, along the continuous lines of the brickwall in Fig.~\ref{fig.1}\cite{bahamon}. The ac field is described by $\omega=2\pi f$ and $F$, the frequency and the field intensity, respectively.

A Floquet-Fourier transformation, considering the so called Floquet eigenstates $|l_1,l_2,m\rangle=|l_1,l_2\rangle\otimes|m\rangle$ with $m$ indicating the photon number states, redefines the Hamiltonian $H$ as an infinite matrix with the elements given by\cite{pablo2004}:
\begin{displaymath} 
\Bigl[(E-m\hbar\omega
-\mathcal{E}_{l_1,l_2})\delta_{l_1^{\prime}l_1}\delta_{l_2^{\prime}l_2}-\Bigl\{{V_x\over2}(\delta_{l_1^{\prime},l_1-1}+\delta_{l_1^{\prime},l_1+1})\delta_{l_2^{\prime}l_2}\Bigr.\Bigr.
\end{displaymath} 
\begin{displaymath}
\Bigl.\Bigl.+{V_y\over2}e^{i2\pi\alpha
l_1}(\delta_{l_2^{\prime},l_2-1}+\delta_{l_2^{\prime},l_2+1})\delta_{l_1^{\prime}l_1}\Bigr\}\Bigr]\delta
_{m'm} 
\end{displaymath} 
\begin{equation}\label{Eq.3}
=\left(
\begin{array}{l}
ea_xFl_1 \\ 
ea_yFl_2
 \end{array}
 \right)\delta_{l_1^{\prime}l_1}\delta_{l_2^{\prime}l_2}(\delta _{m^{\prime},m-1}+\delta
_{m^{\prime},m+1}) \ .
\end{equation}

This matrix is truncated at dimensions given by $L_xL_y(2M+1)$. $L_x$ and $L_y$ are the 
lateral sizes of the graphene finite lattice in number of atomic sites, while $M$ 
is the maximum photon index.  Since the ac field couples the \emph{photon replicas}, for example, the Floquet state indexed by $m$ to states labeled by $m-1$ or $m+1$, multiple photon processes become relevant with increasing field intensity and $M$ settles how many replicas are taken into account. 
This Floquet matrix is a tridiagonal block matrix with 
$L_x\times L_y$ diagonal blocks, given by $\mathsf{E}^M=(E-m\hbar\omega)\mathsf{I}+\mathsf{H_0}$ representing a photon replica with 
the matrix elements given by the left hand side of Eq.~\ref{Eq.3}. The coupling of 
system to the intense ac electric field is represented by the off-diagonal 
blocks ${\mathcal F}$  with the elements given by 
${\mathcal F}=e\left(\begin{array}{l}a_xl_1\\a_yl_2\end{array}\right)F\delta_{l_1^{\prime}l_1}\delta_{l_2^{\prime}l_2}$.
The problem is numerically handled by means of a
renormalization procedure\cite{pablo2004}, based on the associated Green's 
function. The final result is the dressed 
Green's function for one of the photon replicas, say $m=0$, and a 
quasi-density of Floquet states, $\rho({\mathcal E}+i\eta)$ can then be
obtained as
$ 
\rho({\mathcal E}+i\eta)=-\frac{1}{\pi}~\mathrm{Im}\left[~\mathrm{Tr}~G_{MM}~\right],  
$
in the atomic sites basis.

\section{Numerical Results: Landau level replicas and field induced quantum rings}

\begin{figure}[ht]
\includegraphics[scale=0.6]{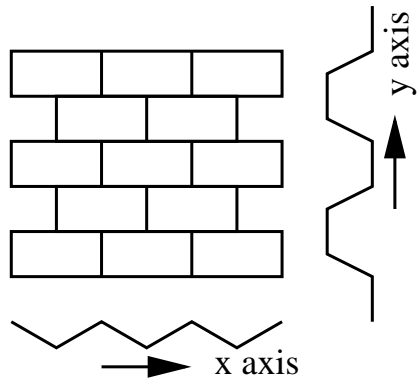}
\caption{The bricklayer model for the graphene lattice. At the right(bottom) side an armchair(zigzag) edge is illustrated in the geometry of the real honeycomb lattice.\label{fig.1}}
\end{figure}

The construction of our system is based on the topological equivalence between the brickwall-like and honeycomb lattices, as already pointed out by 
Iye \emph{et al}\cite{iye}, and as illustrated in Fig.~\ref{fig.1} for a rectangular dot of $7\times6$ carbon-like sites. The actual system discussed in the following  is a rectangular dot of $21\times20$ sites\cite{comment}. The edges of this rectangular graphene dot are zigzag along the $x$ direction with period $a_x$ and armchair along the $y$ with period $a_y$ (see representation in Fig.~\ref{fig.1}).

\begin{figure}[h]
\includegraphics[angle=-90,scale=0.35,clip=true]{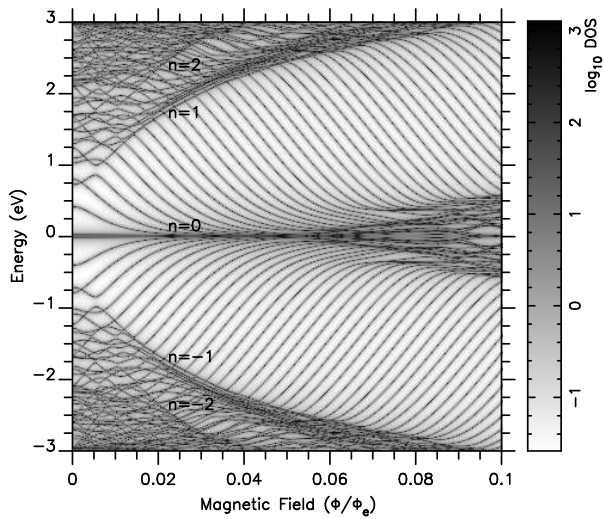}
\caption{The graphene quantum dot spectrum as a function of a perpendicular magnetic flux. Color scale: the darker the line the higher the density of states.\label{fig.2}} 
\end{figure}

The starting point is the inspection of the density of states of the graphene quantum dot in the presence of a perpendicular magnetic field, but in absence of an ac field\cite{bahamon}, shown in Fig.~\ref{fig.2}. In this figure both the 
electron ($E>0$) and hole-like ($E<0$) halfs of the spectrum are shown. Along the horizontal axis at E=0 one can see the central Landau level (LL), $n=0$. At higher(lower) energies one clearly identifies the second, $n=+(-)1$, and the third, $n=+(-)2$, LLs with the typical $\sqrt{nB}$ dependence. Between the central and the second electron(hole)-like LLs ($n=+(-)1$) we see the edge states going down(up) in energy with increasing magnetic field. At the upper(lower) left corner, there is a wide region of strong mixing of the edge and bulk-like states, since there the magnetic length is comparable to the system size, (this is the so called weak field limit, as elegantly discussed previously for a square lattice model\cite{sivan}). 

At higher magnetic fields, the right side of Fig.~\ref{fig.2}, the lattice effects start to develop giving rise to the self-similar Hofstadter spectrum for a honeycomb lattice. Our interest is focused on the energy versus magnetic field window in which the LLs are already well defined but the lattice effects are not yet relevant. Referring to Fig.~\ref{fig.2}, this window of interest spans from $0.02 < \Phi/\Phi_e <0.04$. 
Hence we are focusing on the energy and magnetic field window between the weak field limit and the threshold for the lattice effects. Indeed for the present small systems, bulk-like LLs start to be defined here at very large magnetic fields.  However, the present results can be scaled to lower magnetic fields in larger dots, since the proper length scale for defining a LL is simply a dot dimension larger than the magnetic length.

\begin{figure}[h]
\includegraphics[angle=-90,scale=0.35,clip=true]{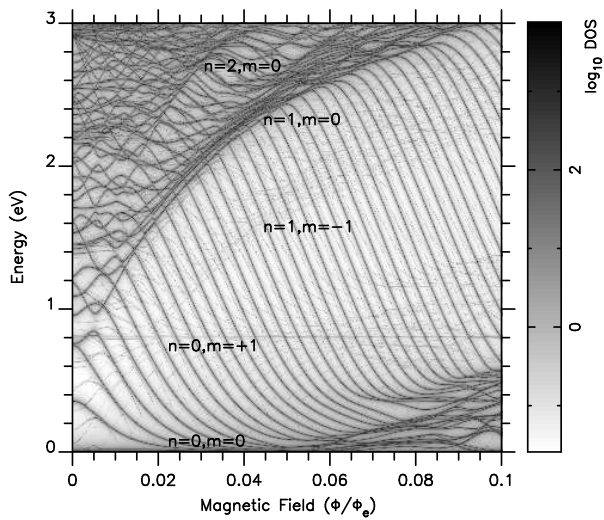} \par
\caption{The same spectrum of Fig. 2 now under an ac field applied parallel to the \textbf{zigzag} direction. The intensity of the field is $0.7\times 10^6$ V/cm  and the photon energy is 0.8 eV ($f=E/h$=193.44 THz).\label{fig.3}}
\end{figure}

The inclusion of an ac field with the electric field component in the graphene plane, parallel to either a zigzag or to an armchair direction is the main discussion stage in what follows. In Fig.~\ref{fig.3}, we show the electron-like part of the quasi density of states of the graphene dot electronic structure dressed by photons, defined by an energy of $\hbar \omega = 0.8 $ eV and a field intensity of $0.7\times 10^6$ V/cm applied parallel to the zigzag direction. Initially these values have to be discussed in order to build up a clear perspective for the present work.  This field intensity corresponds to $eaF\approx 10$ meV, considering $a$ in Eq.~\ref{Eq.3} as the lattice constant for graphene. These are again very high energies and field intensities when compared to actual experimental conditions (the ac field are normally in the terahertz range\cite{vidar} for the effects under scrutiny). Nevertheless, the high absolute values for the ac field intensity and photon energy
  also scale down for larger systems. In summary, the computational limitation to small systems lead the formation of LLs beyond the weak field limit to high magnetic fields and, in order to observe interesting photon dressing effects, the ac field parameters are also increased.  Therefore the discussion of the results has to consider normalized units, $E/\hbar\omega$ and $eaF/\hbar\omega$, leading to a view also valid for realistic and experimentally available parameter sets, as discussed below.

Reports on the microwave induced zero resistance states lead to estimates of the available field intensities in the 100 GHz range of $eaF/\hbar\omega \approx 0.2$. Such a ratio already points toward the necessity of non-perturbative approaches\cite{pablo2004}. Hence, in Fig.~\ref{fig.3}  we limit ourselves to even lower intensities $eaF/\hbar\omega <0.1$, in order to carefully follow the modification induced by dressing with photons the bare electronic structure of graphene. In the spectrum shown in Fig.~\ref{fig.3} we clearly see the $m=1$ replica of the $n=0$ LL (the horizontal line at $E=\hbar\omega=0.8$ eV) while only faint signatures of the $m=-1$ replica of  the $n=1$ LL can be perceived (leading to a shadowy region below $n=1$ LL), as well as replicas of edge states.

\begin{figure}[h]
\includegraphics[angle=-90,scale=0.35,clip=true]{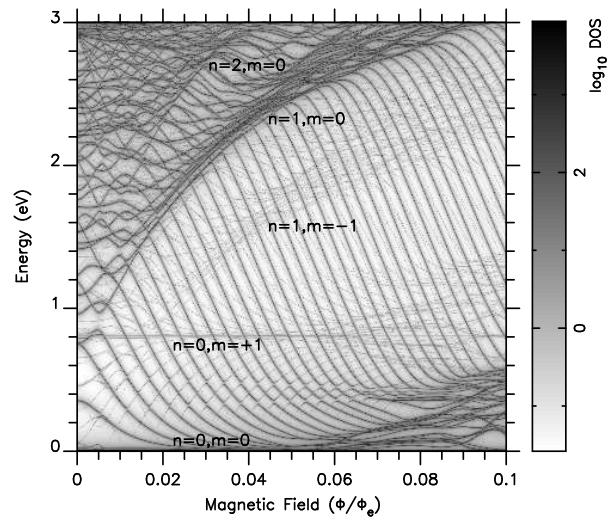}
\caption{Same as Fig.\ref{fig.3} but with the ac field parallel to an \textbf{armchair} direction.\label{fig.4}}
\end{figure}

Fig.~\ref{fig.4} depicts the quasi density of states for an ac field with the same frequency and intensity, but now with the electric field polarized along the armchair direction. Albeit the same field intensity, the replicas seem to be more intense, in particular the edge states ones, as can be particularly seen in an energy stripe around $0.4$ eV. This structure in the set of quasi edge states will be addressed in the following by means of further increasing the amplitude of the ac field.
So far these results indicate a clear difference between the intensity of the photon replicas for fields applied either in the zig-zag and armchair directions, but a clear description of the effect is still missing.

\begin{figure}[h]
\includegraphics[angle=-90,scale=0.35,clip=true]{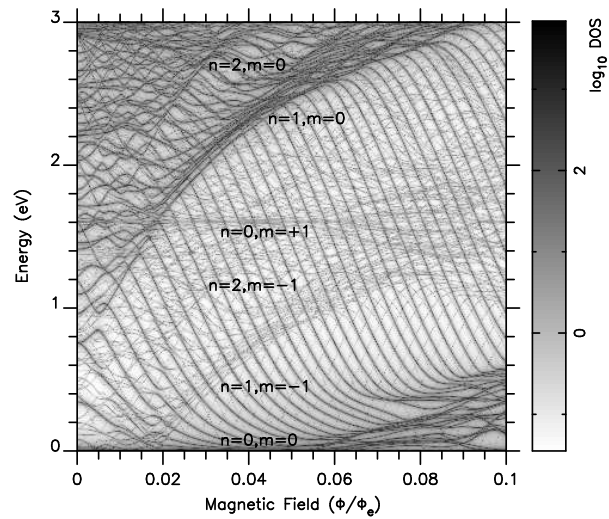}
 \caption{A spectrum of a graphene quantum dot under an ac field intensity of $3.5\times 10^6$ V/cm ($eaF=50$ meV) applied in the \textbf{zigzag} direction of the lattice. Here the photon energy is 1.6 eV (386.88 THz).\label{fig.5}}
\end{figure}

In this context, the quasi density of states for the graphene dot under an ac field along the zigzag direction is shown for higher frequency and field intensity parameters in Fig.~\ref{fig.5}: $\hbar \omega = 1.6$ eV and  $eaF/\hbar\omega \approx 0.03$. By comparing with Fig. 3 and having in mind the photon doubled frequency, two replicas of bulk LLs can be seen: now, besides the $m=-1$ of LL $n=1$, an analogous replica for LL $n=2$ can also be seen. A very strong replica of the central LL can be seen at $E=1.6$ eV, which now merges into  the weak field limit states. 

\begin{figure}[h]
\includegraphics[angle=-90,scale=0.35,clip=true]{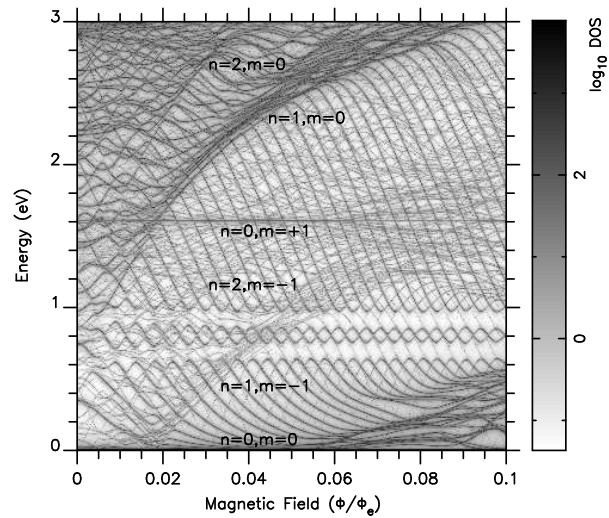}
\caption{The same as Fig.~\ref{fig.5} with the ac field parallel to an \textbf{armchair} direction. We observe an ac field induced quantum ring spectrum of the coupled edge states.\label{fig.6}} 
\end{figure}

Besides the anisotropy in the intensity of the replicas that could be experimentally identified, no clear coupling between quasi states belonging to different replicas have been observed for this case of ac electric field parallel to the zigzag direction. Indeed, anticrossings between quasi states are expected in several systems leading to interesting effects like collapsing superlattice minibands due to intense ac field induced dynamic localization\cite{holthaus} or, more recently and in the context of the present work, new gaps openings in graphene systems \cite{aoki}. Such couplings between different quasi states are revealed by increasing the intensity of an ac field polarized along the armchair direction, Fig.~\ref{fig.6}. Now the hole-like edge states associated to the $m=1$ replica of the central LL (rising in energy with increasing magnetic field) anticross with the electron-like edge states associated to the bare ($m=0$) LL (lowering in energy with increasing magnetic field). Therefore, these anticrossings occur at half the energy separation of both replicas, namely at $\hbar \omega/2 = 0.8$ eV  for the chosen parameters. This picture can be clearly identified by the $gedanken imaging$ of superposing hole-like edge states shown in the lower half of the bare spectrum in Fig. 2 with the electron-like edge states of the very same bare spectrum. Such superposition, actually impossible in a bare system, becomes feasible with the building up of the LL replicas induced by the intense ac field. The appropriate anticrossing behavior leading to the formation of a nicely defined band is connected to the proper symmetry of an ac field along the armchair direction. 

A careful inspection of the miniband build up by the anticrossings of hole-like and electron-like states reveals a periodic modulation with increasing magnetic field. The period of this modulation, $\Delta\Phi/\Phi_e \approx 0.004$, represents a flux quantum through an area corresponding approximately 250 graphene unit cells, i.e., of the order of the dot size. This suggests that this miniband shows the behavior of a quantum ring spectrum\cite{bahamon,beenakker} near the edges of the dot. The simplest model for a quantum ring is a one dimensional tight-binding ring of sites, enclosing a magnetic flux, which can actually be treated analytically\cite{tan,kotlyar}. In experimental systems, like a nanostructured two dimensional electronic gas in GaAs, quantum rings have finite thickness but bona fide quantum ring spectrum can be identified with these very miniband modulations with magnetic field\cite{fhurer}.

One should have in mind that such periodic modulation of the electronic structure is an outcome of the ring geometry, which is absent in our system. The actual potential felt by the electron is a photon dressed potential with a different effective symmetry\cite{pawel}, here the case of a ring. From another point of view, the formation of a quantum ring like electronic structure without a ring structure but in the presence of an ac field is a unique feature of a graphene quantum dot. An ac field induced quasi quantum ring is only possible due to the hole-like edge states associated to the central LL $m=1$ replica anticrossing with the $m=0$ electron-like edge states. It is worth to mention that it could be seen as an emergent property of the Dirac-like electronic structure, since such an effect cannot be observed for conventional two dimensional electronic systems for which below the lowest LL there is only a magnetic barrier\cite{sivan} and all LLs show a linear dispersion with magnetic field.

\section{Discussion: the quest for the field polarization effect}

A remaining important question concerns the field polarization dependence. At a first glance, like mentioned in the introduction, the consequences of the intrinsic anisotropy of grapheme could be explored particularly in a low field quantum Hall regime, due to the insensitivity of microwave induced zero resistance states to the field polarization in GaAs based systems. 

Here qualitatively different pictures emerge with different ac field orientations. The quantum ring like spectrum only appears for the ac electric field parallel to the armchair direction. Nevertheless, the field induced LL replicas appear irrespective of the field polarization, albeit a quantitative difference in the intensity. Since the quantum ring like miniband is build up by means of proper anticrossings, rather involved symmetry properties should be involved.

A further hint refers to the interaction between bulk LLs and their replicas. Looking to the region of the spectrum in Fig.6 where the $m=1$ replica of the $n=0$ LL crosses with the bulk ($m=0$) $n=1$ LL, we see that this crossing occurs very near the weak field limit. Therefore, if we do not consider high photon energies, the investigation of the coupling between these LLs is hindered by the presence of edge states. It is also due to the small size of the dot considered here (restricted by numerical costs) that we have to consider very high photon energies to follow the interactions between bulk LLs and their replicas. In spite of these features, the results suggest that the $m=1,n=0$ LL replica  seems to preserve its identity throughout the entire range down to the zero magnetic field limit, a very different situation from what is expected for LLs with either a $\sqrt{nB}$ or linear dispersion with the magnetic field\cite{pablo2004}. However, a closer observation reveals th
 at this preservation of the identity of the LL replicas occurs only for the armchair case. For ac fields parallel to the zigzag direction, Fig. 5, the bulk LL merges with the whispering gallery of edge states in the weak field limit and is completely smeared out. An analogous smearing out happens to replicas of LLs with linear dispersion with the magnetic field, observed for square lattice models \cite{pablo2004}. Therefore, electric fields parallel to the zigzag direction seem to have similar properties than electric fields parallel to one of the sides of a square lattice. It should be mentioned that real material edges should be a mixture of zigzag and armchair edges and the robustness of the effect has to be addressed in further work on such ac field induced quantum rings.  In actual geometrically defined quantum rings, the characteristic magnetic field periodic spectrum is robust considering edge disorder \cite{bahamon}. 

\section{Final remarks}

The several results depicted here may be summarized in two main findings. (i) The unique electronic structure of graphene near the Dirac point, in the presence of a magnetic field, leads to anisotropic response to the ac field orientation, as revealed in the quasi density of states.  (ii) The ac field induces a quasi-quantum ring (only for the ac field along the armchair direction) within the edge states spectrum, an emergent effect due to the flatness of the central LL and the special coupling among electronic-like and hole-like replicas of edge states. 
\section{Acknowledgement}

The authors would like to acknowledge the financial support from CNPq in the framework of a Latin American research network. ALCP also acknowledges support from FAPESP, PAS and PHR, an additional partial support from CNPq and VRAcad-UNMSM, respectively.


\begin{thebibliography}{100}
\bibitem{comment} The present system is rather small ($\approx 400$ sites), while the smallest experimental graphene quantum dots are of the order of 20~000 sites. Nevertheless, the features of the spectra discussed here can be scaled up to larger systems, as discussed in the text.
\end{thebibliography}
\end{document}